\documentclass[conference]{IEEEtran}
\usepackage{cite}
\usepackage{amsmath,amssymb,amsfonts}
\usepackage{algorithmic}
\usepackage{graphicx}
\usepackage{textcomp}
\usepackage{xcolor}
\usepackage{hyperref}
\usepackage{lipsum}

\def\BibTeX{{\rm B\kern-.05em{\sc i\kern-.025em b}\kern-.08em
    T\kern-.1667em\lower.7ex\hbox{E}\kern-.125emX}}
\begin{document}

\title{Overview and Performance Analysis of Various Waveforms in High Mobility Scenarios}

\author{
	\IEEEauthorblockN{Yu Zhou, Haoran Yin, Jiaojiao Xiong, Shiyu Song, Jiajun Zhu, Jinming Du, Haibo Chen, and Yanqun Tang$^*$}
	\IEEEauthorblockA{\textit{School of Electronics and Communication Engineering, Sun Yat-sen University, China}}
	\IEEEauthorblockA{Email: \{zhouy633, yinhr6, xiongjj7, songshy7, zhujj59, dujm3\}@mail2.sysu.edu.cn, *tangyq8@mail.sysu.edu.cn}
}

\maketitle

\begin{abstract}
In the high-mobility scenarios of next-generation wireless communication systems (beyond 5G/6G), the performance of orthogonal frequency division multiplexing (OFDM) deteriorates drastically due to the loss of orthogonality between the subcarriers caused by large Doppler frequency shifts.
Various emerging waveforms have been proposed for fast time-varying channels with excellent results.
 In this paper, we classify these waveforms from the perspective of their modulation domain and establish a unified framework to provide a comprehensive system structure comparison.
Then we analyze bit error rate (BER) performance of each waveform in doubly selective channels.
Through the discussions on their complexity and compatibility with OFDM systems, we finally give the candidate waveform suggestions.

\end{abstract}


\section{Introduction}
One of the most important scenarios of the next-generation wireless systems and standards (beyond 5G/6G) is extremely high mobility (e.g., high-speed railway systems, vehicle-to-infrastructure, and vehicle-to-vehicle). 
It is well known that orthogonal frequency division multiplexing (OFDM), which has already been adopted in the 5G cellular systems, achieves near-optimal performance in time-invariant  frequency selective channels. 
However, due to the large Doppler frequency shifts in high-mobility scenarios, orthogonality between different subcarriers in the OFDM system is broken, leading to a drastic performance degradation \cite{b1}.

Therefore, many new waveforms are investigated for fast time-varying channels recently. 
Orthogonal chirp division multiplexing (OCDM) based on the discrete Fresnel transform (DFnT) outperforms OFDM in terms of bit error rate (BER) in multipath channels \cite{b2}, \cite{b3}. 
Nevertheless, OCDM cannot achieve full diversity in general time-varying channels, since its diversity order depends on the delay-Doppler profile of the channel.
While affine frequency division multiplexing (AFDM) based on the discrete affine Fourier transform (DAFT), can achieve the full diversity. 
Unlike DFnT, DAFT is a generalization of the discrete Fourier transform (DFT) and is characterized by two parameters that can be adapted to better cope with the doubly selective channels \cite{b4,b5,b6,b01,b03,b02}.

In addition to the chirp-based multicarrier waveforms, the orthogonal time frequency space (OTFS) and orthogonal delay-Doppler division multiplexing (ODDM) were proposed in the delay-Doppler (DD) domain \cite{b7,b8,b11}. 
Although OTFS shows superior performance to OFDM in time-varying channels \cite{b8}, \cite{b10},
OTFS does not have its own orthogonal transmission pulse in the DD domain as ODDM.
Thus, ODDM outperforms  OTFS in terms of out-of-band emission (OOBE) and BER by achieving perfect coupling between the modulated signal and the DD channels \cite{b11}. 

Apart from designing waveforms from the conventional time-frequency (TF) domain and the DD domain, some modulation waveforms implemented in other domains are proposed. Orthogonal time sequency multiplexing (OTSM) and orthogonal delay scale space (ODSS) multiplex information symbols in the delay-sequency domain and the delay-scale domain, respectively \cite{b12,b13,b14}. The underlying transforms of OTSM and ODSS are the Walsh Hadamard transform (WHT), and the discrete Mellin transform (DMT). In particular, OTSM offers similar BER to OTFS but with lower complexity \cite{b13}. ODSS has better BER performance compared to OTFS and OFDM in wideband time-varying channels \cite{b14}.

To get pertinent candidate waveform suggestions for next-generation wireless communications, it is essential to compare the performance of various new waveforms more clearly. 
First of all, we analyze the correlations between them from the perspective of their modulation domain and then obtain a unified framework by deeply discussing their system models. 
After that, we focus on the analysis and comparison of the BER performance of each waveform. 
The simulated results demonstrate our previous judgment about the intrinsic difference of these waveforms.
 Finally, based on the preceding analyses and demonstrations, we further discuss the development prospect of our aforementioned waveforms and give the candidate waveform suggestions.


\textit{Notations}: The following notations will be followed in this paper: a, \textbf{a}, \textbf{A} represent a scalar, vector, and matrix, respectively. (·)$ ^T $ is the transpose operator, (·)$ ^H $ is the Hermitian transpose operation, $\otimes$ is the Kronecker product, $ \operatorname{vec}(\boldsymbol{A}) $ is the column-wise vectorization of the matrix $ \boldsymbol{A} $, $ \operatorname{vec}_{M,N}^{-1}(\boldsymbol{r}) $ is the matrix formed by folding a vector $ \boldsymbol{r} $ into a \textit{N×M} matrix by filling it column wise, and $ \textbf{I} $ denotes the identity metrics.
\begin{figure*}[htbp]
	\centerline{\includegraphics[width=1\textwidth]{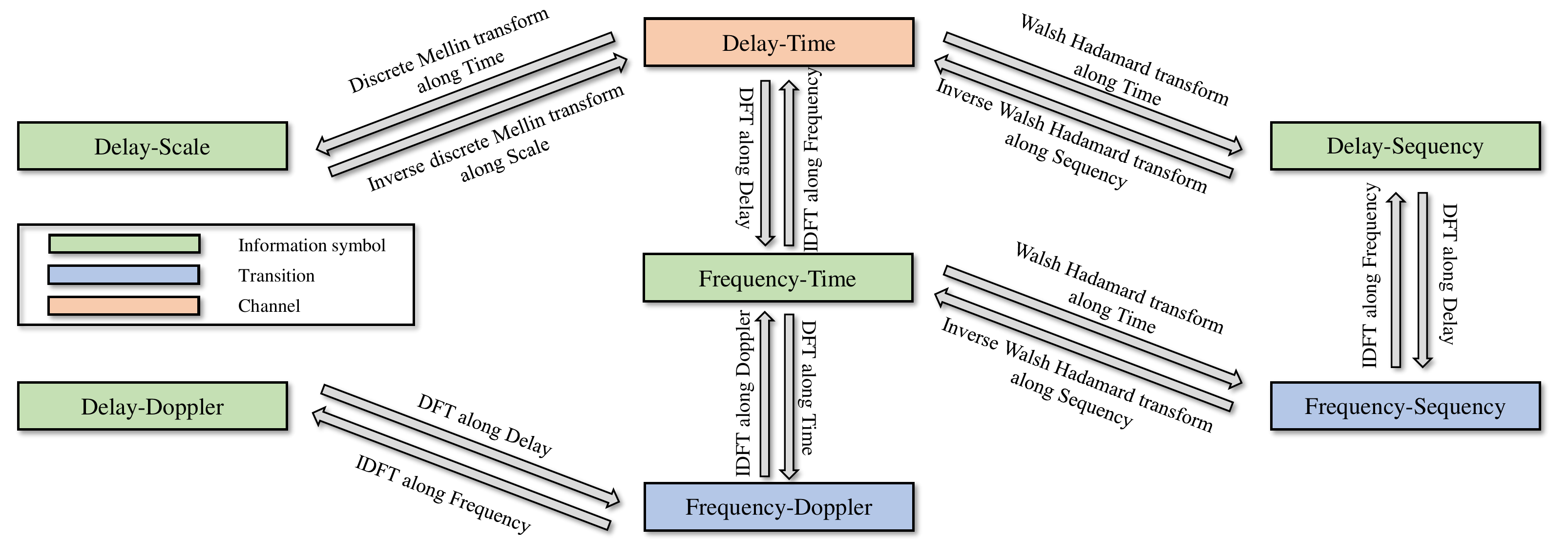}}
	\caption{Transformation relations between different domains along six dimensions (time, frequency, delay, Doppler, scale and sequency) with six transforms (DFT/IDFT, DMT/IDMT and WHT/IWHT).}
	\label{fig2}
\end{figure*}

\section{Waveforms Overview and System Model}\label{s3}
In this section, we provide the overview and system model of the aforementioned waveforms including OFDM, OCDM, OTFS, AFDM, OTSM, ODDM, and ODSS. 
We classify them according to their modulation domain, i.e., time-frequency, delay-Doppler, delay-sequency, and delay-scale domain. 
As we know, the essence of waveform designs is to adopt some kind of mathematical transformation so that the information symbols can achieve orthogonality between different subchannels in the corresponding domain.

The transformation relations among different domains are shown in Fig. \ref{fig2}. 
We can observe that there are three colors corresponding to three types of seven domains, i.e., information symbol, channel and transition.
Each domain can be transformed through special mathematical transforms. 
For instance, OTFS multiplexes information symbols in DD domain, and transfers information symbols from DD domain to TF domain via inverse symplectic finite Fourier transform (ISFFT), i.e., DFT along the delay axis and inverse DFT (IDFT) along the Doppler axis. 
Finally, TF domain is converted to delay-time domain via IDFT along the frequency axis, and the signal is transmitted in delay-time domain.
At the same time, we can see that channel plays an important role in the above-mentioned transformation relations.

A doubly selective channel model with $P$ propagation paths can be written as
\begin{equation}\label{eq1}
	h(\tau, \nu)=\sum_{i=1}^{P} h_{i} \delta\left(\tau-\tau_{i}\right) \delta\left(\nu-\nu_{i}\right),
\end{equation}
where $\delta(\cdot)$ is the Dirac delta function, $h_i$, $\nu_i$ and $\tau_i$ are the complex gain, Doppler shift, and the delay shift associated with the $ i $-th path, respectively. Let $\tau_{max}$ and $\nu_{max}$ denote the maximum delay spread and Doppler spread of the
doubly selective channel, respectively. Thus we have $0 \leq \tau_{i} \leq \tau_{\max }$ and $-\nu_{\max } \leq \nu_{i} \leq \nu_{\max }$.

The corresponding continuous time-varying channel impulse response function can be written as
\begin{equation}\label{eq2}
g(\tau, t)=\int_{\nu} h(\tau, \nu) \mathrm{e}^{j 2 \pi \nu(t-\tau)} d \nu.
\end{equation}

In order to obtain the principles of waveform design adapting to the doubly selective channels, we establish a unified framework based on the classified domains, which includes the transceiver and channel as shown in Fig. \ref{fig1} . 
From Fig. \ref{fig1}, we can easily differ the mathematical transforms for modulation and demodulation of seven waveforms.
Through this framework we can also understand the complexity and compatibility with OFDM-based systems. 

\subsection{Time-Frequency domain}
OFDM can be seen as a chirp-based modulation waveform with a linearly varying instantaneous frequency of zero. OCDM and AFDM are both based on chirp modulation waveforms. Therefore, we consider them as chirp-based waveforms and discuss them in time-frequency domain.
\subsubsection{OFDM}
In the case of OFDM, data symbols $\mathbf{X}$ can be written as
\begin{equation}
\mathbf{X}=[\mathbf{X}(\mathbf{0}), \mathbf{X}(\mathbf{1}), \ldots, \mathbf{X}(\mathbf{M}-\mathbf{1})]^T,
\end{equation}
where $\mathbf{X}(m)=[X(m, 0), X(m, 1), \ldots, X(m, N-1)]^T $.


As shown in Fig. \ref{fig1}, $X(n, m)$ is the data symbol transmitted on the $m$-th subcarrier of the $n$-th OFDM symbol. The data symbols $\mathbf{X}=[X(0), X(1), \ldots, X(N-1)]^T$ transmitted on each $m$-th OFDM symbol are fed into the IDFT block, and the resulting signal in time domain can be written as
\begin{equation}\label{eq7}
	s[n]=\frac{1}{\sqrt{N}} \sum_{m=0}^{N-1} X[m] e^{j\frac{2 \pi}{N} n m},
\end{equation}

In matrix form, \eqref{eq7} can be rewritten as
\begin{equation}
\mathbf{s}=\mathbf{F}^H \mathbf{X},
\end{equation}
where $\mathbf{F}$ is the DFT matrix with entries $ e^{-j 2 \pi m n / N}/\sqrt{N} $.
After the addition of the cyclic prefix (CP), the signal is transmitted into the channels. 

\subsubsection{OCDM}
Let $ \boldsymbol{x} $ denote an \textit{N}×1 vector of quadrature amplitude modulation (QAM) symbols. After the serial to parallel operation, \textit{N}-points inverse DFnT (IDFnT) is performed to map $ \boldsymbol{x} $ to the time domain as


\begin{equation}\label{1}
s(n)=\frac{1}{\sqrt{N}} e^{j \frac{\pi}{4}} \sum_{m=0}^{N-1} x[m]\times\left\{\begin{array}{ll}
		e^{-j \frac{\pi}{N}(n-m)^{2}} \quad N \equiv 0(\bmod 2) \\
		e^{-j \frac{\pi}{N}\left(n-m-\frac{1}{2}\right)^{2}}  N \equiv 1(\bmod 2).
	\end{array}\right.
\end{equation}
	
The DFnT matrix can be decomposed using the equation $ \boldsymbol{\Phi}=\boldsymbol{\Theta_{2}}\boldsymbol{F}\boldsymbol{\Theta_{1}} $, where $ \boldsymbol{\Theta_{1}} $ and $ \boldsymbol{\Theta_{2}} $ are two diagonal matrices given by
\begin{equation}
\Theta_{1}(m)=e^{-j \frac{\pi}{4}} \times\left\{\begin{array}{ll}
	e^{j \frac{\pi}{N} m^{2}} & N \equiv 0(\bmod 2) \\
	e^{j \frac{\pi}{4 N}} e^{j \frac{\pi}{N}\left(m^{2}+m\right)} & N \equiv 1(\bmod 2)
\end{array}\right.
\end{equation}
and
\begin{equation}
\Theta_{2}(n)=\left\{\begin{array}{ll}
	e^{j \frac{\pi}{N} n^{2}} & N \equiv 0(\bmod 2) \\
	e^{j \frac{\pi}{N}\left(n^{2}-n\right)} & N \equiv 1(\bmod 2).
\end{array}\right.
\end{equation}

\eqref{1} can be rewritten in matrix form as 
\begin{equation}
\mathbf{s}=\boldsymbol{\Theta_{1}}^{H} \mathbf{F}^{H} \boldsymbol{\Theta_{2}}^{H} \boldsymbol{x}=\boldsymbol{\Phi}^{H}\boldsymbol{x}.
\end{equation}

After the transmission over the channel, discarding CP and performing \textit{N}-points DFnT, the received sample matrix  can be presented in the matrix form as
\begin{equation}
	\boldsymbol{y}=\boldsymbol{\Theta_{2}} \mathbf{F} \boldsymbol{\Theta_{1}}\textbf{H}\boldsymbol{\Theta_{1}}^{H} \mathbf{F}^{H} \boldsymbol{\Theta_{2}}^{H}\boldsymbol{x}+\tilde{\boldsymbol{w}}=\textbf{H}_{\mathrm{eff}} \boldsymbol{x}+\tilde{\boldsymbol{w}},
\end{equation}
where $ \tilde{\boldsymbol{w}} \sim \mathcal{C N}\left(0, \sigma_{c}^{2} \textbf{I}\right) $  is an additive Gaussian noise vector and $ \textbf{H} $ is the matrix representation of the communications channel in the time domain.

As shown in Fig. \ref{fig1}, the DFnT can be implemented by DFT in three
steps:
\begin{itemize}
	\item
	multiplying the chirp phase $ \boldsymbol{\Theta_{1}} $,
	\item
	performing the DFT,
	\item
	multiplying the other chirp phase $ \boldsymbol{\Theta_{2}} $,
\end{itemize}
where $ \boldsymbol{\Theta_{1}} $ and $ \boldsymbol{\Theta_{2}} $ are diagonal matrices whose \textit{m}-th diagonal
entries are $ \boldsymbol{\Theta_{1}}(m) $ and $ \boldsymbol{\Theta_{2}}(m) $, respectively.

\subsubsection{AFDM}\label{afdm}
Let $ \boldsymbol{x} $ denote an \textit{N}×1 vector of QAM symbols. After the serial to parallel operation, \textit{N} points inverse DAFT (IDAFT) is performed to map $ \boldsymbol{x} $ to the time domain as
\begin{equation}\label{eq3}
s[n]=\frac{1}{\sqrt{N}} \sum_{m=0}^{N-1} x[m] e^{j 2 \pi\left(c_{1} n^{2}+\frac{1}{N} m n+c_{2} m^{2}\right)},
\end{equation}
where $c_1$ and $ c_2 $ are the AFDM parameters, and $ n =
0, \ldots , N-1 $. Then, a chirp-periodic prefix (CPP) is added with a length of $ L_{cp} $, which is any integer greater than or equal to the value in samples of the maximum delay spread of the channel. The prefix is
\begin{equation}
s[n]=s[N+n] e^{-j 2 \pi c_{1}\left(N^{2}+2 N n\right)}, \quad n=-L_{\mathrm{cp}}, \cdots,-1.
\end{equation}
We can notice that CPP equals CP whenever $ 2Nc_1 $ is an
integer and \textit{N} is even.

\eqref{eq3} can be rewritten in matrix form as
\begin{equation}\label{eq4}
	\mathbf{s}=\boldsymbol{\Lambda}_{c_{1}}^{H} \mathbf{F}^{H} \boldsymbol{\Lambda}_{c_{2}}^{H} \boldsymbol{x},
\end{equation}
where 
\begin{equation}\label{eq5}
\boldsymbol{\Lambda}_{c_i}=\operatorname{diag}\left(e^{-j 2 \pi c_i n^{2}}, n=0,1, \ldots, N-1\right).
\end{equation}

After the transmission over the channel, discarding CPP and performing \textit{N}-points DAFT, the received sample matrix  can be written in the matrix form as
\begin{equation}
\boldsymbol{y}=\textbf{H}_{\mathrm{eff}} \boldsymbol{x}+\tilde{\boldsymbol{w}},
\end{equation}
where $ \tilde{\boldsymbol{w}} \sim \mathcal{C N}\left(0, \sigma_{c}^{2} \textbf{I}\right) $  is an additive Gaussian noise vector and $ \textbf{H}_{\mathrm{eff}}=\boldsymbol{\Lambda}_{c_{2}} \textbf{F} \boldsymbol{\Lambda}_{c_{1}} \textbf{H} \boldsymbol{\Lambda}_{c_{1}}^{\mathrm{H}} \textbf{F}^{\mathrm{H}} \boldsymbol{\Lambda}_{c_{2}}^{\mathrm{H}} $, $ \textbf{H} $ being the matrix representation of the communications channel in the time domain.

\begin{figure*}[htbp]
	\centerline{\includegraphics[width=1\textwidth]{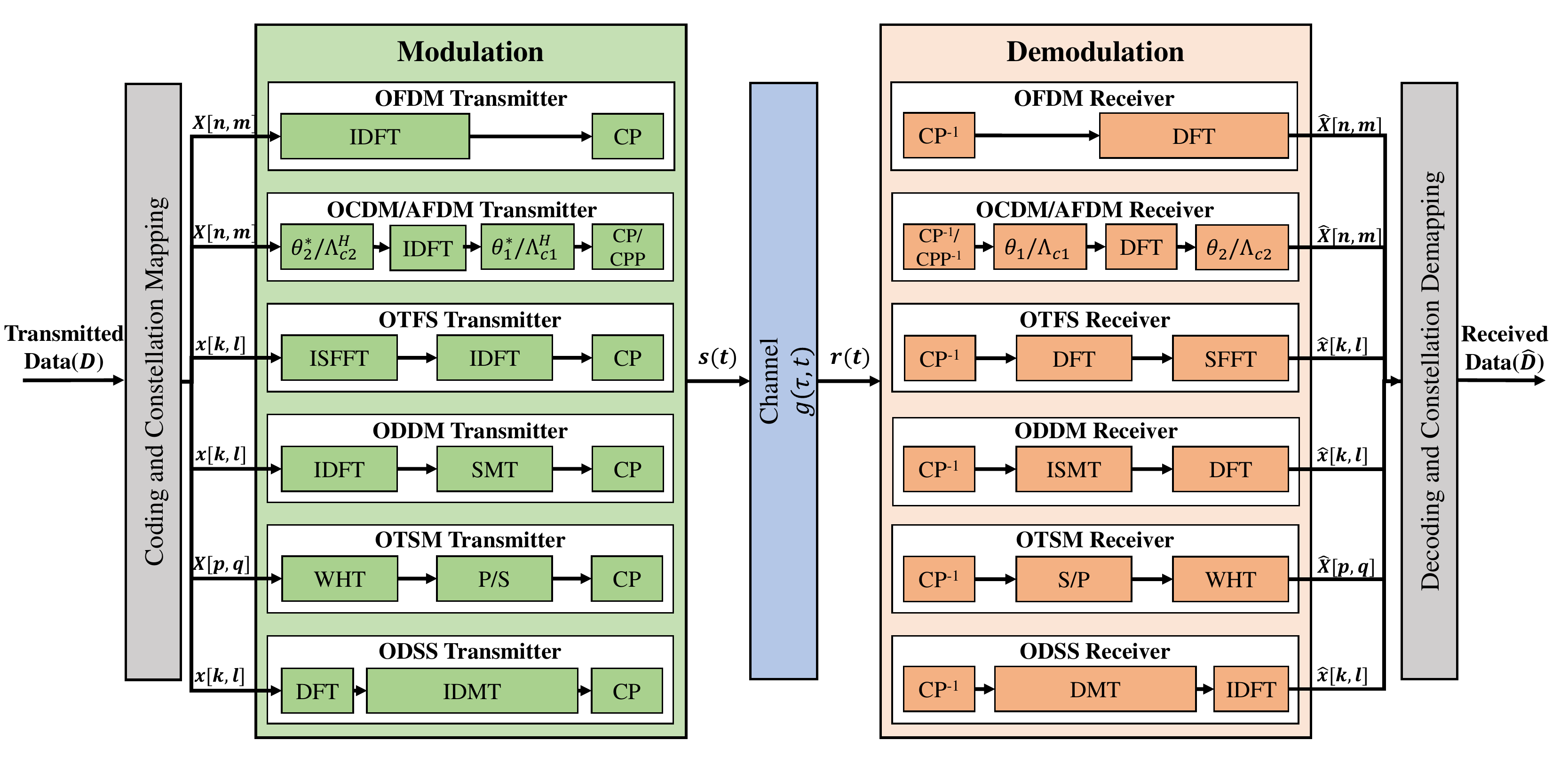}}
	\caption{Unified framework for seven waveforms, i.e., OFDM, OCDM, AFDM, OTFS, ODDM, OTSM and ODSS.}
	\label{fig1}
\end{figure*}

As shown in Fig. \ref{fig1}, the DAFT can be implemented by DFT in three steps:
\begin{itemize}
	\item
	multiplying the chirp phase $ \boldsymbol{\Lambda}_{c_{1}} $,
	\item
	performing the DFT,
	\item
	multiplying the other chirp phase $ \boldsymbol{\Lambda}_{c_{2}} $,
\end{itemize}
where $ \boldsymbol{\Lambda}_{c_{1}} $ and $ \boldsymbol{\Lambda}_{c_{2}} $ are diagonal matrices.

It is worth to note that DFT and DFnT are two special cases of DAFT. 
When $ c_1 $ and $ c_2 $ are both zero or $ \frac{1}{2N} $, DAFT becomes DFT and DFnT, respectively.

\subsection{Delay-Doppler domain}
\subsubsection{OTFS}
The block diagram of OTFS modulation and demodulation is shown in Fig. \ref{fig1}. The information bits to be transmitted, after bit-to-symbol mapping, are multiplexed onto a discrete two-dimension (2D) DD domain grid with a size of $\hat{N} \times \hat{M}$. The information symbols, $ x[k,l] $, are mapped from the discrete DD domain to the TF domain, $ X[n,m] $, by the ISFFT as follows:
\begin{equation}
X[n, m]=\frac{1}{\hat{N} \hat{M}} \sum_{l=0}^{\hat{M}-1} \sum_{k=0}^{\hat{N}-1} x[k, l] e^{j 2 \pi\left(\frac{n l}{\hat{N}}-\frac{m k}{\hat{M}}\right)},
\end{equation}
where $m \in\{0,1, \ldots, \hat{M}-1\}, n \in\{0,1, \ldots, \hat{N}-1\}$. 

Heisenberg transform converts the 2D TF data, $X[n, m]$, to a 1D continuous time-series, $s(t)$, given by
\begin{equation}\label{eq10}
s(t)=\sum_{m=0}^{\hat{M}-1} \sum_{n=0}^{\hat{N}-1} X[n, m] e^{j 2 \pi n \Delta f_{\rm OTFS}(t-m T)} g_{\mathrm{tx}}(t-m T),
\end{equation}
where $g_{\mathrm{tx}}(t)$ is the transmit pulse shaping function, $T$ and $ \Delta f_{\rm OTFS} $ represent the time and frequency domain sample
intervals of the time-frequency plane.
%

\subsubsection{ODDM}
ODDM transmits symbols $ x[k,l] $ on a discrete 2D delay-Doppler domain grid of size $\hat{N} \times \hat{M}$, as shown in Fig. \ref{fig1}. The CP-free waveform of the DD plane multicarrier modulation is given by
\begin{equation}\label{eq8}
s(t)=\sum_{k=0}^{\hat{M}-1} \sum_{l=0}^{\hat{N}-1} x[k, l] \check{g}_{t x}\left(t-k \frac{T}{\hat{M}}\right) e^{j 2 \pi \frac{l}{\hat{N} T}\left(t-k \frac{T}{\hat{M}}\right)},
\end{equation}
where $ \check{g}_{t x}(t) $ is the transmit pulse. In particular, a DD plane orthogonal pulse  $ \check{g}_{t x}(t) $ is adopted.

ODDM symbols are sampled at a rate of $ \frac{1}{T} $, and staggered at an interval of $ \frac{T}{\hat{M}} $. Just like OFDM, ODDM uses the IDFT to obtain the time-domain
discrete samples of the $ k $-th ODDM symbol as
\begin{equation}
s[k, \dot{l}]=\sum_{l=0}^{\hat{N}-1} x[k, l] e^{j 2 \pi \frac{\dot{l}l}{\hat{N}}}, \dot{l}=0, \ldots \hat{N}-1,
\end{equation}
where $ \dot{l} $ represents the index of the time-domain discrete samples spaced by $ T $.

To stagger $ \hat{M} $ ODDM symbols at an interval of $ \frac{T}{\hat{M}} $, the aforementioned $ N $ time-domain discrete
samples need to be upsampled by $ \hat{M} $ to obtain $ \hat{M}\hat{N} $ discrete
samples, given by
\begin{equation}
	\begin{array}{r}
		\mathbf{s}[k]=[\overbrace{0, \ldots, 0}^m, s[k, 0], \overbrace{0, \ldots, 0}^{\hat{M}-1}, s[k, 1], \overbrace{0, \ldots, 0}^{\hat{M}-1}, \\
		\ldots, \overbrace{0, \ldots, 0, s[k, \hat{N}-1]}^{\hat{M}-1}, \overbrace{0, \ldots, 0}^{\hat{M}-m-1}].
	\end{array}
\end{equation}

After the parallel to serial (P/S) operation, and pulse shaping, the signal $ s(t) $ in time domain is transmitted in the channel. The upsampling, P/S and pules shaping operation are named staggered multitone (SMT) modulation, as shown in Fig. \ref{fig1}.

Assuming $ a(t) $ present a time-symmetric real-valued square-root Nyquist pulse, where $ \int_{-\infty}^{+\infty}|a(t)|^{2} d t= \frac{1}{\hat{N}}$.
Based on the pulse shaping method for ODDM modulation proposed in \cite{b11}, \eqref{eq8} can be rewritten as
\begin{equation}\label{eq9}
	s(t)=\sum_{k=0}^{\hat{M}-1} \sum_{\dot{l}=0}^{\hat{N}-1} \sum_{l=0}^{\hat{N}-1} x[k, l] e^{j 2 \pi \frac{\dot{l}l}{\hat{N}}} a\left(t-k \frac{T}{\hat{M}}-\dot{l} T\right).
\end{equation}

The transmit pulse of ODDM  $ u(t) $ can be denoted as \cite{b11}
\begin{equation}
u(t)=\sum_{\dot{l}=0}^{\hat{N}-1} a(t-\dot{l} T).
\end{equation}
Therefore, \eqref{eq9} can be rewritten as
\begin{equation}\label{eq11}
s(t)=\sum_{k=0}^{\hat{M}-1} \sum_{l=0}^{\hat{N}-1} x[k, l] u\left(t-k \frac{T}{\hat{M}}\right) e^{j 2 \pi \frac{l}{\hat{N} T}\left(t-k \frac{T}{\hat{M}}\right)}.
\end{equation}

Finally, considering the prepended CP, the definition of $u(t)$ can be extended to
\begin{equation}
u_{c p}(t)=\sum_{\dot{l}=-1}^{\hat{N}-1} a(t-\dot{l} T) .
\end{equation}
Then, the CP-included ODDM waveform becomes
\begin{equation}
s_{c p}(t)=\sum_{k=0}^{\hat{M}-1} \sum_{l=0}^{\hat{N}-1} x[k,l] u_{c p}\left(t-k \frac{T}{\hat{M}}\right) e^{j 2 \pi \frac{l}{\hat{N} T}\left(t-k \frac{T}{\hat{M}}\right)}.
\end{equation}
By comparing \eqref{eq10} with \eqref{eq11}, we can observe that the transmit pulse in OTFS modulation is a rectangle pulse as a special case for pulse shaping in ODDM modulation.

\subsection{Delay-Sequency domain}
\textbf{\textit{OTSM: }}OTSM is a single-carrier modulation scheme, which offers similar BER to OTFS. The information symbols are multiplexed in the delay-sequency domain using WHT. Note that sequency is the number of zero-crossings per unit interval. Since WHT does not require multiplicative operations and requires only addition and subtraction operations, the OTSM modulation/demodulation complexity is significantly low as compared to OFDM and OTFS modulation \cite{b12}, \cite{b13}.

Let $ \boldsymbol{x}, \boldsymbol{y} \in C^{\hat{N} \hat{M} \times 1} $ be the transmitted and received information symbols. The modulation and demodulation system model of OTSM is shown in Fig. \ref{fig1}. At the transmitter, the information symbols $ \boldsymbol{x}=\left[\boldsymbol{x}_{0}^{\mathrm{T}}, \cdots, \boldsymbol{x}_{\hat{M}-1}^{\mathrm{T}}\right]^{\mathrm{T}} $ are split into vectors $ \boldsymbol{x}_{m} \in \mathbb{C}^{\hat{N} \times 1}, m=0, \ldots, \hat{M}-1 $. The symbol vectors
are arranged into a delay-sequency matrix $ \mathbf{X} \in \mathbb{C}^{\hat{M} \times \hat{N}} $
\begin{equation}
\mathbf{X}=\left[\boldsymbol{x}_{0}, \boldsymbol{x}_{1}, \ldots, \boldsymbol{x}_{\hat{M}-1}\right]^{\mathrm{T}},
\end{equation}
where the matrix column and row indices represent the delay and sequency indices of the delay-sequency grid, respectively. Then, a $ \hat{N} $-point inverse WHT (IWHT) is applied on each of these symbol vectors (rows) to transform it to the delay-time domain.
\begin{equation}
\tilde{\mathbf{X}}=\left[\tilde{\boldsymbol{x}}_{0}, \tilde{\boldsymbol{x}}_{1}, \ldots, \tilde{\boldsymbol{x}}_{\hat{M}-1}\right]^{\mathrm{T}}=\mathbf{X} \mathbf{W}_{\hat{N}},
\end{equation}
where $ \mathbf{W}_{\hat{N}} $ is the normalized $ \hat{N} $-point WHT matrix.

The matrix $ \tilde{\mathbf{X}} $ contains the delay-time samples which are column-wise vectorized to obtain the
time-domain samples $ \mathbf{s} \in \mathbb{C}^{\hat{N} \hat{M} \times 1} $ to be transmitted into the physical channel
\begin{equation}\label{2}
\mathbf{s}=\operatorname{vec}(\tilde{\mathbf{X}}).
\end{equation}
The transmitter operation \eqref{2} can be expressed in the
simple matrix form as
\begin{equation}\label{3}
	\mathbf{s}=\mathbf{P} \left(\mathbf{I}_{\hat{M}} \otimes \mathbf{W}_{\hat{N}}\right) \boldsymbol{x},
\end{equation}
where $ \mathbf{P} $ is the row-column interleaver matrix.
The transmitter operation in \eqref{3} can be simplified as
\begin{equation}
	\mathbf{s}=\left(\mathbf{W}_{\hat{N}} \otimes \mathbf{I}_{\hat{M}}\right) (\mathbf{P} \boldsymbol{x}).
\end{equation}

A CP of length $l_{\max }$ is added to the time-domain samples, which are pulse shaped, digital-to-analog converted and transmitted into the wireless channel as $s(t)$.

At the receiver, the received time-domain signal $ r(t) $ is processed via analog-to-digital conversion and CP removal, yielding time-domain vector $\mathbf{r} \in \mathbb{C}^{\hat{N} \hat{M} \times 1}$. The received time domain samples $\mathbf{r}$ are folded into the matrix $\tilde{\mathbf{Y}}$ column-wise as
\begin{equation}
\tilde{\mathbf{Y}}=\left[\tilde{\boldsymbol{y}}_0, \tilde{\boldsymbol{y}}_1, \ldots, \tilde{\boldsymbol{y}}_{\hat{M}-1}\right]^{\mathrm{T}}=\operatorname{vec}_{\hat{M}, \hat{N}}^{-1}(\mathbf{r}).
\end{equation}

The received delay-sequency information symbols are obtained by taking an $ \hat{N} $-point WHT of the rows of received delay-time matrix $ \tilde{\mathbf{Y}} $ as
\begin{equation}
\mathbf{Y}=\left[\boldsymbol{y}_0, \boldsymbol{y}_1, \ldots, \boldsymbol{y}_{\hat{M}-1}\right]^{\mathrm{T}}=\tilde{\mathbf{Y}} \mathbf{W}_{\hat{N}}.
\end{equation}

The receiver operation can be rewritten in matrix form as
\begin{equation}
\boldsymbol{y}=\left(\mathbf{I}_{\hat{M}} \otimes \mathbf{W}_{\hat{N}}\right) \left(\mathbf{P}^{\mathrm{T}} \mathbf{r}\right),
\end{equation}
where $\boldsymbol{y}=\left[\boldsymbol{y}_0^{\mathrm{T}}, \cdots, \boldsymbol{y}_{\hat{M}-1}^{\mathrm{T}}\right]^{\mathrm{T}}$.

\subsection{Delay-Scale domain}
\textbf{\textit{ODSS: }}The underlying Mellin transform of ODSS enjoys a scale-invariance property , i.e., the Mellin transform of the signal $ \sqrt{a} x(a \alpha), a>0, \alpha>0 $, is same as that of the original signal, $ x(\alpha) $, except for a phase shift. The Mellin transform of a signal $ x(\alpha), \alpha>0 $, is defined by
\begin{equation}
\mathcal{M}_{x}(\beta) \triangleq \int_{0}^{\infty} \frac{1}{\sqrt{\alpha}} x(\alpha) e^{j 2 \pi \beta \log (\alpha)} d \alpha,
\end{equation}
where $ \alpha $ is the scale variable, and $ \beta \in \mathbb{R} $ is the Mellin variable. The inverse Mellin transform is given by
\begin{equation}
	x(\alpha) \triangleq \frac{1}{\sqrt{\alpha}} \int_{-\infty}^{\infty} \mathcal{M}_x(\beta) e^{-j 2 \pi \beta \log (\alpha)} d \beta, \alpha>0.
\end{equation}

The information bits are multiplexed onto the discrete 2D Mellin-Fourier domain of size, $M_{\text {tot }}=\sum_{n=0}^{\hat{N}-1} M(n)$, where $ q $ is the ratio of the geometric sampling in the scale domain, and $M(n)=\left\lfloor q^n\right\rfloor$. ODSS maps the data symbols, $\{x[k, l]: k=0,1, \ldots, \hat{N}-1, l=0,1, \ldots, M(k)\}$, in the
discrete Mellin-Fourier space to the 2D sequence, $X[n, m]$, in the delay-scale domain by taking an inverse discrete Mellin transform (IDMT) along the scale axis and a discrete Fourier transform along the delay axis, as follows:
\begin{equation}
	X[n, m]=\frac{q^{-n / 2}}{\hat{N}} \sum_{k=0}^{\hat{N}-1} \frac{\sum_{l=0}^{M(k)-1} x[k, l] e^{j 2 \pi\left(\frac{m l}{M(k)}-\frac{n k}{\hat{N}}\right)}}{M(k)}.
\end{equation}

The ODSS modulator converts the 2D TF data, $X[n, m]$, to a 1D continuous time-series, $s(t)$, given by
\begin{equation}
s(t)=\sum_{n=0}^{\hat{N}-1} \sum_{m=0}^{M(n)-1} X[n, m] q^{n / 2} g_{\mathrm{tx}}\left(q^n\left(t-\frac{m}{q^n W}\right)\right),
\end{equation}
where $ g_{tx}(t) $ is the transmit pulse shaping function , and $W$ is the signal bandwidth.

\section{Simulation Results}\label{s4}
In this section, we provide some simulation results to assess the performance of the aforementioned waveforms. The complex gains $ h_i $ are generated as independent complex Gaussian random variables with zero mean and $ 1/P $ variance. BER values are obtained using $ 10^5 $ different channel realizations.

\begin{figure}[htbp]
	\centerline{\includegraphics[width=0.6\textwidth]{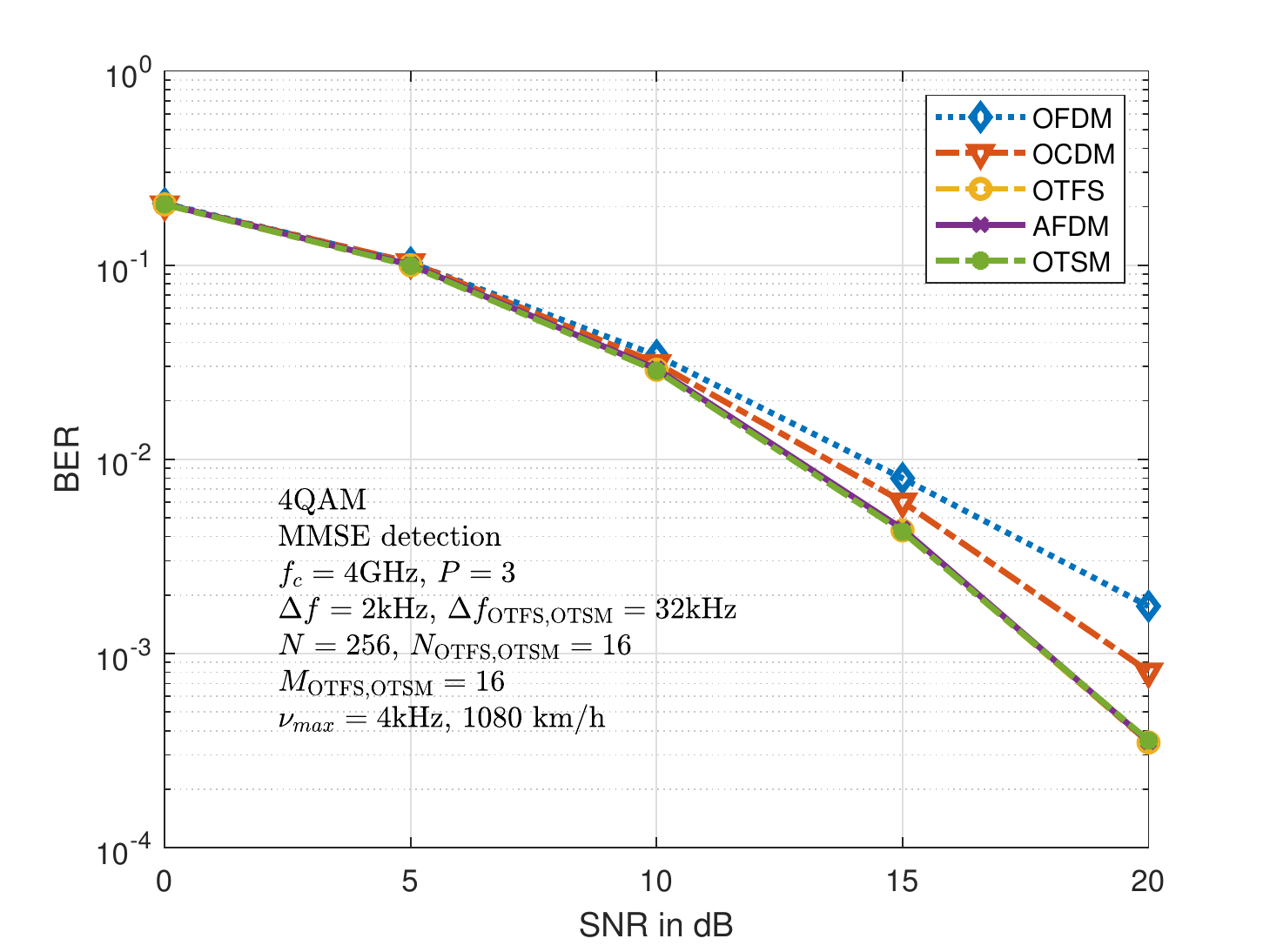}}
	\caption{BER performance of OFDM, OCDM, OTFS, AFDM, and OTSM in a three-path channel using MMSE detection.}
	\label{fig3}
	\vspace{-0.2em}
\end{figure}

Fig. \ref{fig3} shows the simulated BER performance of these waveforms with parameters setting of  $ f_c =4\rm GHz $, $ P=3 $, $ \bigtriangleup f=2\rm kHz $, $ \bigtriangleup f_{\rm OTFS,OTSM}=32\rm kHz $, $ \nu_{max}=4\rm kHz $ (corresponding to a speed of 1080 $ \rm km/h $), $ N=256 $, $ N_{\rm OTFS,OTSM}=16 $, $ M_{\rm OTFS,OTSM}=16 $ to ensure the same resources are occupied by all the considered waveforms, 4QAM and minimum mean square error (MMSE) detection are used.

Firstly, we can observe that the performance of OFDM is the worst in the high-mobility scenarios.
This is mainly due to large Doppler frequency shifts and the loss of orthogonality among different subcarriers, resulting in inter-carrier interference.
Then, OCDM has a better performance than OFDM as its better path separation capabilities.
OTFS outperforms OFDM and OCDM since OTFS operates in the DD coordinate system where all modulated symbols experience the same channel gain, and hence OTFS is able to extract the full channel diversity under a limited signal-to-noise ratio (SNR).
Moreover, we can see that AFDM and OTSM have identical performance with OTFS because both of them can maintain good orthogonality of subcarriers in the corresponding domain to bear information symbols.
Due to limited space, here we only discuss the comparison of ODDM and ODSS while not giving the simulated results.
As described in \citen{b11,b14}, ODDM outperforms the OTFS by achieving perfect coupling between the modulated signal and the DD channel, and ODSS has better performance than OTFS in wideband time-varying channels.

In order to meet the demands of next-generation wireless communications in high-mobility scenarios, it is important to focus on the BER performance.
On the other hand, the waveform selections for 6G need to have good compatibility with the existing communication systems.
Thus, in the sequel, we will emphasize three waveforms (i.e., AFDM, OTFS, OTSM) based on their outstanding BER performance.
From Fig. \ref{fig1}, and Fig. \ref{fig3}, we can contradictorily observe that OTSM has a better BER performance with bad compatibility.
It is known that OTFS and AFDM have good compatibility with the existing OFDM-based systems.
OTFS only needs to add a precoder (ISFFT) at the transmitter of the existing OFDM system.
 AFDM needs to add two matrix-multiply operations ($ \boldsymbol{\Lambda}_{c_{1}} $ and $ \boldsymbol{\Lambda}_{c_{2}} $) before and after the DFT, respectively, which is the definition of DAFT.

In conclusion, with the best BER performance and good compatibility with OFDM, AFDM can satisfy the requirements of high-mobility scenarios in next-generation wireless communication systems.
Besides, in our previous research, we have proved that AFDM is currently the only mathematically proven waveform that can achieve full diversity with multiple antennas under the doubly selective channels (without introducing additional precoding procedures), and have advantages over
OTFS in terms of pilot overhead \cite{b01}, \cite{b03}. 
Furthermore, as mentioned in Section \ref{afdm}, the complexity of modulation/demodulation of AFDM is almost low with OFDM.
In a word, in the points of BER performance, compatibility and complexity, AFDM is the best choice for high-mobility communications in future wireless systems.

\section{Conclusion and Prospect}\label{s5}
The existing OFDM-based waveforms can't supply the physical foundations to service the high-mobility scenarios in next-generation wireless communications. 
The basic reasons are that these waveforms are just designed to match the frequency selective channels while not considering the doubly selective characteristics in fast time-varying channels.
The overviewed waveforms in this paper are all constructed in doubly selective channels, although they are designed in different domains mentioned above.
We establish a unified framework to provide a comprehensive performance comparison. 
Based on the simulated results of BER performance and the analysis of compatibility and complexity, we conclude and give the suggestion of AFDM as the candidate waveform.
The recent research on AFDM modulation will greatly promote the development of physical layer wireless transmission technology.
This waveform will apply to next-generation multiple access (NGMA) for future wireless communications, and be widely used in  integrated sensing and communications (ISAC) scenarios with its high-precision time and frequency resolution.

\end{document}